\documentclass[fleqn,12pt,twoside]{article}
\usepackage{espcrc1}

\usepackage{graphicx}

\newcommand{\AmS}{{\protect\the\textfont2
  A\kern-.1667em\lower.5ex\hbox{M}\kern-.125emS}}

\title{Chemical and mechanical spinodals a unique 
 liquid-gas instability }

\author{Philippe Chomaz\address[MCSD]
{GANIL CEA/DSM - CNRS/IN2P3 BP 5027 F-14076 Caen CEDEX 5, France}%
        \ and
        Jérôme Margueron\addressmark[MCSD]}
       
\begin{document}

\maketitle

\begin{abstract}
We demonstrate that the instabilities of asymmetric nuclear matter at
sub-saturation densities do not present two types of instabilities as
usually discussed but a unique one. The associated order parameter is
everywhere dominated by the isoscalar density and so the transition is of
liquid-gas type even in the so-called chemical instability region.
\end{abstract}

\begin{figure}[tbph]
\center
\includegraphics[scale=0.3]{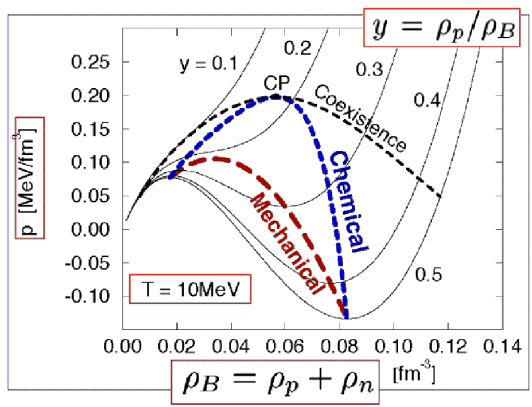}
\caption{The pressure as a function of the total
density for a fixed temperature and different asymmetry. The region where
the isotherms present a back bending are usually related to a mechanical
instability. However, one can define a broader region of instability by
allowing the isospin fluctuation (traditionally discussed as a chemical
instability). We will show that this classification is artificial since it
does not correspond to a modification of the physical properties of the
instabilities. }
\label{fig0}
\end{figure}

Below saturation density, the nuclear interaction is expected to
lead to a liquid-gas phase transition \cite{ber80}. Recently, a converging
ensemble of experimental signals seems to have established the phase
transition. One is the spinodal decomposition \cite{bor01}.  It has been
argued that asymmetric nuclear matter do not only present a mechanical
instability for which the density is the order parameter but also a broader
chemical instability associated with fluctuations of the matter isospin
content  \cite{mul95}.  Indeed, it is usually argued that  it exists a
region in which the compressibility at constant asymmetry and constant
isospin is negative (see figure (\ref{fig0})) leading to the interpretation
that the system is mechanically unstable. Above a maximum instability the
isotherms at constant asymmetry does not presents any back bending leading
to the idea that the system is mechanically stable. However, looking at the
equilibrium of the chemical potentials one can see that above this maximum
asymmetry for mechanical instabilities the system may amplify fluctuations
in the proton neutron concentration leading to a second instability region
usually called chemical instabilities.      

However, we have recently shown that this splitting of the spinodal region
in to two types of instabilities a mechanical and a chemical one is not
correct and that ANM present only one type of instabilities \cite{mar03}.
The associated order parameter is dominated by the isoscalar density and so
the transition is of liquid-gas type. The instability goes in the direction
of a restoration of the isospin symmetry leading to a fractionation
phenomenon. Our conclusions are model independent since they can be related
to the general form of the asymmetry energy. They are illustrated using
density functional approaches derived from Skyrme and Gogny effective
interactions.


Let us consider ANM characterized by a proton and a neutron densities $\rho
_{i}=$ $\rho _{p}$, $\rho _{n}$. These densities can be transformed in a set
of 2 mutually commuting charges $\rho _{i}=$ $\rho _{1}$, $\rho _{3}$ where $%
\rho _{1}$ is the density of baryons, $\rho _{1}=\rho _{n}+\rho _{p},$ and $%
\rho _{3}$ the asymmetry density $\rho _{3}=\rho _{n}-\rho _{p}$. In
infinite matter, the extensivity of the free energy implies that it can be
reduced to a free energy density~: $F(T,V,N_{i})=V\mathcal{F}(T,\rho _{i}).$
The system is stable against separation into two phases if the free energy
of a single phase is lower than the free energy in all two-phases
configurations. This stability criterium implies that the free energy
density is a convex function of the densities $\rho _{i}$. A local necessary
condition is the positivity of the curvature matrix~: 
\begin{equation}
\left[ \mathcal{F}_{ij}\right] =\left[ \frac{\partial ^{2}\mathcal{F}}{%
\partial \rho _{i}\partial \rho _{j}}|_{T}\right] \equiv \left[ \frac{%
\partial \mu _{i}}{\partial \rho _{j}}|_{T}\right]  \label{eq5}
\end{equation}
where we have introduced the chemical potentials $\mu _{j}\equiv \frac{%
\partial F}{\partial N_{j}}|_{T,V,N_{i}}=\frac{\partial \mathcal{F}}{%
\partial \rho _{j}}|_{T,\rho _{i\not{=}j}}$.

\begin{figure}[tbph]
\center
\includegraphics[scale=0.3]{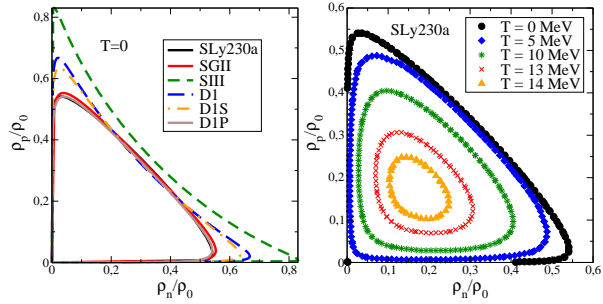}
\caption{Projection of the spinodal contour in the
density plane : left, for Skyrme (SLy230a~\protect\cite{cha97}, SGII~ 
\protect\cite{ngu81}, SIII~\protect\cite{bei75}) and Gogny models (D1~%
\protect\cite{gog75}, D1S~ \protect\cite{ber91}, D1P~\protect\cite{far99}) ;
right, temperature dependence of the spinodal zone computed for the SLy230a
case.}
\label{fig2}
\end{figure}

We show in the left part of Fig.~\ref{fig2} the spinodal contour in ANM for
several forces. It exhibits important differences. In the case of SLy230a
force (as well as SGII, D1P), the total density at which spinodal
instability appears decreases when the asymmetry increases whereas for SIII
(as well as D1, D1S) it increases up to large asymmetry and finally
decreases. We observe that all forces which fulfill the global requirement
that they reproduce symmetric nuclear matter (SNM) equation of state as well
as the pure neutron matter calculations, leads to the same curvature of the
spinodal region. We can appreciate the reduction of the instability when we
go away from SNM. However, large asymmetries are needed to induce a sizable
effect. The temperature dependence of the spinodal contour can be
appreciated in the right part of Fig.~\ref{fig2}. As the temperature
increases the spinodal region shrinks up to the critical temperature for
which it is reduced to SNM critical point. However, up to a rather high
temperature ( $5$ $\mathrm{MeV}$) the spinodal zone remains almost identical
to the zero temperature one.




In the considered two-fluids system, the $[\mathcal{F}_{ij}]$ is a $2*2$
symmetric matrix, so it has 2 real eigenvalues $\lambda ^{\pm}$~\cite{bar01}%
~: 
\begin{equation}
\lambda ^{\pm}=\frac{1}{2}\left( \mathrm{Tr}\left[ \mathcal{F}_{ij}\right]
\pm \sqrt{\mathrm{Tr}\left[ \mathcal{F}_{ij}\right] ^{2}-4\mathrm{Det}\left[ 
\mathcal{F}_{ij}\right] }\right)  \label{eq23}
\end{equation}
associated to eigenvectors $\mathbf{\delta \rho }^{\pm}$ defined by ($i\neq
j $) 
\begin{equation}
\frac{{\delta \rho }_{j}^{\pm}}{{\delta \rho }_{i}^{\pm}}=\frac{\mathcal{F}%
_{ij}}{\lambda ^{\pm}-\mathcal{F}_{jj}}=\frac{\lambda ^{\pm}-\mathcal{F}_{ii}%
}{\mathcal{F}_{ij}}  \label{eq25}
\end{equation}
Eigenvectors associated with negative eigenvalue indicate the direction of
the instability. It defines a local order parameter since it is the
direction along which the phase separation occurs. The eigen values $\lambda 
$ define sound velocities, $c$, by ${c}^{2}=\frac{1}{18m}\rho _{1}\,\lambda
. $ In the spinodal area, the eigen value $\lambda $ is negative, so the
sound velocity, $c$, is purely imaginary and the instability time $\tau $ is
given by $\tau =d/|c|$ where $d$ is a typical size of the density
fluctuation.

The requirement that the local curvature is positive is equivalent to the
requirement that both the trace ($\mathrm{Tr}[\mathcal{F}_{ij}]=\lambda
^{+}+\lambda ^{-})$ and the determinant ($\mathrm{Det}[\mathcal{F}%
_{ij}]=\lambda ^{+}\lambda ^{-})$ are positive 
\begin{equation}
\mathrm{Tr}[\mathcal{F}_{ij}]\geq 0,\hbox{ and }\mathrm{Det}[\mathcal{F}%
_{ij}]\geq 0  \label{eq6}
\end{equation}
The use of the trace and the determinant which are two basis-independent
characteristics of the curvature matrix clearly stresses the fact that the
stability analysis should be independent of the arbitrary choice of the
thermodynamical quantities used to label the state e.g. $(\rho _{p}$, $\rho
_{n})$ or $(\rho _{1}$, $\rho _{3})$.


A different discussion can be found in the literature~\cite{mul95,bao97} and
we will now clarify the relation of this discussion with the eigen modes
analysis. Indeed, from the thermodynamical relation 
\begin{equation}
\rho _{n}\,\mathrm{Det}[\mathcal{F}_{ij}]=\frac{\partial \mu _{p}}{\partial y%
}|_{T,P}\,\,\frac{\partial P}{\partial \rho _{1}}|_{T,y}  \label{eq9}
\end{equation}
one can be tempted to relate separately $\frac{\partial \mu _{p}}{\partial y}%
|_{T,P}$ and $\frac{\partial P}{\partial \rho _{1}}|_{T,y}$ to the two
eigenvalues $\lambda ^{+}$ and $\lambda ^{-}$. The discussion of the sign of 
$\lambda ^{+}$ and $\lambda ^{-}$ is replaced by the discussion of the sign
of $\frac{\partial \mu _{p}}{\partial y}|_{T,P}$ (also called chemical
instability) and $\frac{\partial P}{\partial \rho _{1}}|_{T,y}$ (also called
mechanical instability). This is correct in SNM and the mechanical
instability is associated with density fluctuations whereas the chemical
instability is associated with the isospin density fluctuation. In the case
of ANM, because natural symmetries are lost, theses relations break down
where one can show that~\cite{bar01} 
\begin{eqnarray}
\frac{\partial P}{\partial \rho _{1}}|_{T,y} &=&\frac{\lambda ^{+}}{\sqrt{t}}%
(t\cos \beta +\sin \beta )^{2}+\frac{\lambda ^{-}}{\sqrt{t}}(t\sin \beta
-\cos \beta )^{2} \\
\frac{\partial \mu _{p}}{\partial y}|_{T,P} &=&\rho _{n}\lambda ^{+}\lambda
^{-}\left( \frac{\partial P}{\partial \rho _{1}}|_{T,y}\right) ^{-1}
\end{eqnarray}
where $\beta =1/2\tan \mathcal{F}_{np}/(\mathcal{F}_{pp}-\mathcal{F}_{nn})$
and $t=\rho _{n}N_{0}^{n}/\rho _{p}N_{0}^{p}$. These two equalities
illustrate explicitly that the simplicity of SNM is not preserved in ANM.
Hence, one should come back to eigen analysis of the curvature matrix $[%
\mathcal{F}_{ij}]$. 



The local stability condition in ANM is express by Eq.~\ref{eq6}. If is
violated the system is in the unstable region of a phase transition. Two
cases are then possible~: i) only one eigenvalue is negative and one order
parameter is sufficient to describe the transition or ii) both eigenvalues
are negative and two independent order parameters should be considered
meaning that more than two phases can coexist.

For ANM below saturation density, the case ii) never occurs since the
asymmetry energy has always positive curvature ($\mathcal{F}_{33}$). Indeed,
the asymmetry term in the mass formula behave like $(N-Z)^{2}$ times a
positive function of $A$ showing that the dominant $\rho _{3}$ dependence of
the asymmetry potential energy is essentially quadratic and that $\mathcal{F}%
_{33}$ is a positive function of the total density. Recent Bruckner
calculations in ANM~\cite{vid02} have confirmed the positivity of $\mathcal{F%
}_{33}$. They have parameterized the potential energy with the simple form $%
\mathcal{V}(\rho _{1},\rho _{3})=\mathcal{V}_{0}(\rho _{1})\rho _{1}^{2}+%
\mathcal{V}_{1}(\rho _{1})\rho _{3}^{2}$ with $\mathcal{V}_{0}/\mathcal{V}%
_{1}\sim -3$ and $\mathcal{V}_{0}<0$. This is also true for effective forces
such as Skyrme forces. For example, the simplest interaction with a constant
attraction, $t_{0}$, and a repulsive part, $t_{3}\rho _{1}$, leads to $%
\mathcal{V}(\rho _{1},\rho _{3})=(3\rho _{1}^{2}-\rho _{3}^{2})\beta (\rho
_{1})$ with $\beta (\rho _{1})=(t_{0}+t_{3}/6\rho _{1})/8$. The function $%
\beta (\rho _{1})$ is negative below saturation density, hence the
contribution of the interaction to $\mathcal{F}_{33}$ in the low density
region is always positive.

These arguments show that, below saturation density, the $\rho _{3}$
curvature, $\mathcal{F}_{33}$, is expected to be positive for all
asymmetries. Since the curvature in any direction, $\mathcal{F}_{ii}$,
should be between the two eigenvalues $\lambda ^{-}\leq \mathcal{F}_{ii}\leq
\lambda ^{+}$ we immediately see that if $\mathcal{F}_{33}$ is positive one
eigen curvature at least should remain positive. 
In fact for all models we have studied $\mathcal{F}_{33}$ appears to be
always large enough so that the trace is always positive demonstrating that $%
\lambda ^{+}>0$. Since $\mathrm{Tr}[\mathcal{F}_{ij}]=\mathcal{F}_{nn}+%
\mathcal{F}_{pp}$, this can be related to the positivity of the Landau
parameter $\mathcal{F}_{nn}$ and $\mathcal{F}_{pp}$.

\smallskip The large positive value of $\mathcal{F}_{33}$ also indicates
that the instability should remain far from the $\rho _{3}$ direction i.e.
it should involve total density variation and indeed we will see that in all
models and for all asymmetry the instability direction hardly deviates from
a constant asymmetry direction $(\delta \rho _{3}\ll \delta \rho _{1})$.
This isoscalar nature of the instability can be understood by looking at the
expression of the eigen-modes in the $(\rho _{n};\rho _{p})$ coordinates.
Since $\lambda ^{-}\leq \mathcal{F}_{ii}\leq \lambda ^{+}$, the differences $%
\lambda ^{-}-\mathcal{F}_{ii}$ is always negative demonstrating, using Eq.~%
\ref{eq25}, that the instability is of isoscalar type, if the Landau
parameter $\mathcal{F}_{np}$ is negative. Hence, there is a close link
between the isoscalar nature of the instability and the attraction of the
proton-neutron interaction~\cite{bar01}.

\begin{figure}[htbp]
\center
\includegraphics[scale=0.45]{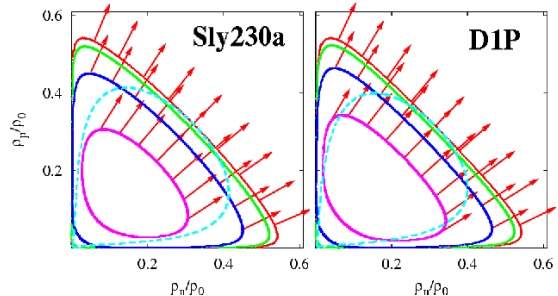}
\caption{This is the projection of the iso-eigen values on the density plane
for Slya (left) and D1P (right). The arrows indicate the direction of
instability. The mechanical instability is also indicated (dotted line).}
\label{fig3}
\end{figure}

Various contour of equal imaginary sound velocity are represented in Fig.~%
\ref{fig3} for SLy230b and D1P interactions. The more internal curve
correspond to the sound velocity $i0.09c$, after comes $i0.06c$, $i0.03c$
and finally 0, the spinodal boarder. For these two recent forces which takes
into account pure neutron matter constraints, the predicted instability
domains are rather similar. We observe that in almost all the spinodal
region the sound velocity is larger than $0.06c$.

\smallskip Let us now focus on the direction of the instability. If $\mathbf{%
\delta \rho }^{-}$ is along $y$=cst then the instability does not change the
proton fraction. For symmetry reasons pure isoscalar $(\delta \rho _{3}=0)$
and isovector $(\delta \rho _{1}=0)$ modes appears only for SNM so it is
interesting to introduce a generalization of isoscalar-like and
isovector-like modes by considering if the protons and neutrons move in
phase ($\delta \rho _{n}^{-}\delta \rho _{p}^{-}>0$) or out of phase ($%
\delta \rho _{n}^{-}\delta \rho _{p}^{-}<0$). Fig.~\ref{fig3} shows the
direction of instabilities along the spinodal boarder and some
iso-instability lines. We observed that instability is always almost along
the $\rho _{1}$ axis meaning that it is dominated by total density
fluctuations even for large asymmetries. 
In fact, the instability direction is always between
the $y$=cst line and the $\rho _{1}$ direction. This shows that the unstable
direction is dominated by isoscalar modes
as expected from the attractive interaction
between proton-neutron. The total density is therefore the dominant
contribution to the order parameter showing that the transition is between
two phases having different densities (i.e. liquid-gas phase transition).
The angle with the $\rho _{1}$ axis is almost constant along a constant $y$
line. This means that as the matter enters in the spinodal zone and then
dives into it, there are no dramatic change in the instability direction
which remains essentially a density fluctuation. Moreover, the unstable
eigenvector drives the dense phase (i.e. the liquid) towards a more
symmetric point in the density plane. By particle conservation, the gas
phase will be more asymmetric leading to the fractionation phenomenon. Those
results are in agreement with recent calculation for ANM~\cite{bar01} and
nuclei~\cite{col02}.


We want to stress that those qualitative conclusions are very robust and
have been reached for all the Skyrme and Gogny forces we have tested (SGII,
SkM$^{*}$, RATP, D1, D1S, D1P,...) including the most recent one (SLy230a,
D1P) as well as the original one (like SIII, D1).


In this paper, we have shown that ANM does not present two types of spinodal
instabilities, a mechanical and chemical, but only one which is dominantly
of isoscalar nature as a consequence of the negativity of the Landau
parameter $\mathcal{F}_{np}$. This general property can be linked to the
positivity of the symmetry energy curvature $\mathcal{F}_{33}$. This means
that the instability is always dominated by density fluctuations and so can
be interpreted as a liquid-gas separation. The instabilities tend to restore
the isospin symmetry for the dense phase (liquid) leading to the
fractionation of ANM. We have shown that changing the asymmetry up to $\rho
_{p}<3\rho _{n}$ does not change quantitatively the density at which
instability appears, neither the imaginary sound velocity compared to those
obtained in SNM. All the above results are not qualitatively modified by the
temperature which mainly introduce a reduction of the spinodal region up to
the SNM critical point where it vanishes. The quantitative predictions
concerning the shape of the spinodal zone as well as the instability times
depends upon the chosen interaction but converge for the various forces
already constrained to reproduce the pure neutron matter calculation.

\end{document}